# Crowdsourcing: a new tool for policy-making?

By

Araz Taeihagh

# Crowdsourcing: a new tool for policy-making?

Araz Taeihagh[1]

Singapore Management University

**Abstract** – Crowdsourcing is rapidly evolving and applied in situations where ideas, labour, opinion or expertise of large groups of people are used. Crowdsourcing is now used in various policy-making initiatives; however, this use has usually focused on open collaboration platforms and specific stages of the policy process, such as agenda-setting and policy evaluations. Other forms of crowdsourcing have been neglected in policy-making, with a few exceptions. This article examines crowdsourcing as a tool for policy-making, and explores the nuances of the technology and its use and implications for different stages of the policy process. The article addresses questions surrounding the role of crowdsourcing and whether it can be considered as a policy tool or as a technological enabler and investigates the current trends and future directions of crowdsourcing.

## Introduction

Crowdsourcing is becoming ubiquitous! In the words of Lehdonvirta and Bright (2015, p. 263): 'If elections were invented today, they would be called Crowdsourcing the Government.' Crowdsourcing (Howe 2006, 2008; Brabham 2008) is rapidly evolving, and is now loosely applied to instances where a relatively large number of individuals are engaged by organisations for their ideas, expertise, opinions, or labour (Lehdonvirta and Bright 2015; Prpić and Shukla 2016). Crowdsourcing has now expanded from focusing on consumers and businesses to non-commercial domains. Crowdsourcing can also increase transparency and broaden citizen engagement in policy-making, and foster citizen empowerment (Fischer 1993; Aitamurto 2012, 2016b; Prpić, Taeihagh and Melton 2015; Liu 2017a). Crowdsourcing has now been employed in policy-making in areas such as urban planning, state and federal policy (Seltzer and Mahmoudi 2013; Aitamurto *et al.* 2016), transportation (Nash 2009), law reform (Aitamurto 2016a) and global governance (Gellers 2016). Furthermore, it has been demonstrated that crowdsourcing has the potential to help address some of the prevailing challenges in data and judgment acquisition for policy design and analysis (Prpić, Taeihagh and Melton, 2014; Taeihagh 2017b).

---

[1] School of Social Sciences, Singapore Management University, 90 Stamford Road, Level 4, Singapore 178903, Singapore. E-mail: araz.taeihagh@new.oxon.org

Despite the recent advances in the use of crowdsourcing in the public sector, only a handful of studies methodologically examine its use in the policy cycle. It has been demonstrated that, although increasing, the use of crowdsourcing in the policy cycle is still limited, and not all of its potential has been realised (Prpić, Taeihagh and Melton, 2015). Scholars have mainly used Open Collaboration (OC) platforms in agenda-setting, problem definition and policy evaluation stages; with a few exceptions, other approaches, such as Tournament Crowdsourcing (TC) or Virtual Labour Markets (VLM), have been neglected.

In the next section, we briefly introduce the concept of crowdsourcing and distinguish between its different general types. We then systematically examine different roles that different types of crowdsourcing can take in the policy cycle and highlight their nuances. We develop a taxonomy of the major types of crowdsourcing to facilitate future studies, distinguishing between procedural or substantive policy tools and front- or back-end policy tools, and take steps to help develop more empirical studies to better understand the efficacy of the use of crowdsourcing in the policy cycle. We then examine the current trends and future direction of crowdsourcing before the concluding remarks.

**Crowdsourcing**

Crowdsourcing is an umbrella term, and the definition and scope of it vary among scholars. Crowdsourcing is used when the dispersed knowledge of individuals and groups are leveraged to take advantage of bottom-up crowd-derived inputs and processes with efficient top-down engagement from organisations through IT, to solve problems, complete tasks, or generate ideas (Howe 2006; 2008; Brabham 2008; 2013a). In the context of public policy this increased access to dispersed knowledge of crowds can enhance knowledge utilisation and learning that can increase the chance of policy success (Bennett and Howlett 1992).

Crowdsourcing can be done in a closed environment, in which 'propriety crowds' are utilised through in-house platforms by an organisation, or carried out using third-party platform crowdsourcing that provides the IT infrastructure and their crowd of participants to the potential pool for organisations to tap into as a paid service (Bayus 2013, Prpić, Taeihagh and Melton, 2015).

In this article, we focus on the three main types of crowdsourcing identified in the literature and try to develop a more nuanced understanding of the crowdsourcing concept and how it applies to the policy cycle (Estellés-Arolas and González-Ladrón-de-Guevara 2012; de Vreede, *et al.*, 2013; Prpić, Taeihagh and Melton 2015).[2] These three general forms of crowdsourcing focus on:

a) microtasking in VLMs (Prpić, Taeihagh and Melton 2014; Luz, Silva and Novais 2015; De Winter *et al.* 2015);
b) TC competition (Schweitzer *et al.* 2012; Zhang *et al.* 2015; Glaeser *et al.* 2016); and
c) OC over the web and social media (Budhathoki and Haythornthwaite 2013; Michel, Gil and Hauder, 2015; Mergel 2015).

VLMs

---

[2] These categorisations are not exclusive or exhaustive, but useful for considering the different roles crowdsourcing can take in the policy cycle. For a review of the state-of-the-art in crowdsourcing, see Prpić (2016).

A VLM is an IT-mediated market that enables individuals to engage in spot labour through conducting microtasks offered by organisations, exemplifying the production model of crowdsourcing in exchange for money (Brabham 2008; Horton and Chilton 2010; Paolacci, Chandler, and Ipeirotis 2010; Prpić, Taeihagh and Melton 2014; Luz, Silva and Novais 2015; De Winter *et al*. 2015).

Microtasks are best known to be offered by Amazon Mechanical Turk (Mturk.com) and Crowdflower (crowdflower.com). They include tasks such as document translation, content moderation, transcription, sentiment analysis, photo and video tagging, and data entry and categorisation (Narula *et al*. 2011, Crowdflower 2016). Such tasks can be broken down into different steps (microtasks) that can be carried out at scale and in parallel by individuals through human computational power.

At the moment, these microtasks are better performed by human computation and through collective intelligence rather than by using computational approaches and artificial intelligence (Taeihagh 2017b). The majority of the microtasks offered on these platforms are repetitive and require low to medium levels of skill, and thus the compensations per task are low, and the labourers involved in the VLM platforms are employed anonymously.[3] In VLM platforms often labourers cannot form teams or groups, and there is only an episodic engagement among them and the platform. This is purely a function of the design of the VLM platforms and can (and will probably) change in future which will enable completion of more sophisticated tasks and more complex interactions among crowds.

TC

In TC, or Idea Competition (Piller and Walcher 2006; Jeppesen and Lakhani 2010; Schweitzer *et al*. 2012; Glaeser *et al*. 2016), organisations post their problems to specialised IT-mediated platforms (Eyreka or Kaggle) or in-house platforms (Challenge.gov: Brabham 2013b). Here, organisers form a competition through the IT-mediated platform and set conditions and rules for the competition, and winner(s) prize. To be considered for the prize, which can range from a few hundred dollars to hundreds of thousands of dollars, individuals or groups (depending on the capabilities of the IT platform and the rules of the contest) post their solutions to the posted problems on the appropriate platform.[4]

---

[3] With respect to their offline identities. However, researchers such as Lease *et al*. (2013) have previously demonstrated that significant amount of information can be exposed about the workers through the VLM websites.

TC platforms mainly aim to attract and maintain more specialised crowds that are interested in a particular area. This can range from open government and innovation (The White House 2010) to computer or data science (Lakhani *et al*. 2010; Taieb and Hyndman 2014). TC platforms attract smaller and more specialised crowds that are capable of solving more complex tasks, and at times choose not to be anonymous to gain reputational benefits from their successful participations (Prpić, Taeihagh and Melton, 2015).

OCs

In OC crowdsourcing, crowds voluntarily engage with the problems /opportunities posted by organisations through IT platforms without expectation of monetary compensation (Crump 2011; Michel, Gil and Hauder, 2015). Starting wikis, and employing online communities and social media to amass contributions, or using project hosting websites such as GitHub for collaboration are examples of OCs (Jackson and Klobas 2013; Crowley *et al*. 2014; Rogstadius *et al*. 2013; Budhathoki and Haythornthwaite 2013; Mergel, 2015; Longo and Kelley 2016).

The level of the crowd's engagement depends on many factors, such the effectiveness of the 'open call', the reach and level of engagement of the IT-mediation platform used by the organisation, and the crowd capital of the organisation (Prpić Taeihagh Melton 2015;

---

[4] https://www.kaggle.com/competitions.

Prpić and Shukla 2013). As an example, as of 30 June 2016, Twitter has more than 313 million monthly active users;[5] however, this does not necessarily translate into significant engagement from the active users of a platform. Numerous factors influence the level of traction, diffusion and ultimately success of an open call in an OC platform. A small number of these factors include the level of prior engagement and popularity of the organisation on the platform, the number of followers and shares of content/calls made by the organisation, and the popularity and stature of the crowds they engage (e.g. attention from celebrities, Nobel laureates), alongside the quality of the content posted (Cha *et al*. 2010; Taeihagh 2017a). Any number of these individuals engaging in the open call can alter, hijack or amplify the agenda of the organisation with their networks (Prpić and Shukla 2013; Prpić, Taeihagh and Melton 2015).

The three principal types crowdsourcing described above have different levels of accessibility, crowd magnitude, crowd specialisations, anonymity, and IT structure, as well as platform framework and interactions (Prpić, Taeihagh and Melton 2015; Taeihagh 2017a: see Table 1). Table 1 demonstrates that different types of crowdsourcing each have unique sets of characteristics, while sharing similarities with other types.

Table 1 Comparison of different types of crowdsourcing (based on Prpić, Taeihagh, Melton 2015 and Taeihagh 2017a)

|  | **Accessibility** | **Crowd magnitude** | **Nature of the crowd** | **Anonymity** | **Platform architecture** | **IT structure** | **Platform interactions** |
|---|---|---|---|---|---|---|---|
| **VLMs** (e.g. Amazon Mturk) | Private | Millions | General | High | Community building and infrastructure provision | Episodic | Information, currency, and virtual services |
| **TC** (e.g. Kaggle) | Private | Hundreds of thousands | Specialised | Medium | Community building | Episodic | Information, currency, and virtual services |
| **OC** (e.g. Twitter) | Public | Hundreds of millions | General | Variable | Community building | Collaborative | Information |

**Crowdsourcing as a policy tool**

Given the brief description of principal types of crowdsourcing, we now examine crowdsourcing as a policy tool using Hood's NATO model (Hood 1986; 2007; Hood and Margetts 2007). In the NATO model, the following four types of resources can be used by governments to address policy problems (see Table 2):

- informational advantage through centrality in various networks (nodality);
- legal power to command, regulate, or delegate (authority);
- financial means, such as the ability to fund or demand taxes (treasure); and
- deploying resources to form organisations and markets, provide goods and services (organisation).

---
[5] https://about.twitter.com/company.

The NATO model does not demand the strict singular dependence of an instrument on one of the four resources. Instead, instruments are categorised according to the primary means they require for successfully addressing their goals. A second distinction used by Hood in characterising various tools is whether they are used for detecting changes in the environment (detector) or for affecting the outside world (effector). Similar to the effector/detector distinction, Howlett (2000) introduced the positive/negative distinction between policy instruments based on whether they encourage or discourage actor participation in the policy process. Another relevant distinction is whether these policy instruments are substantive (directly providing or altering aspects of provision, distribution or delivery of goods and services to the public or governments) or procedural (rather than directly affecting the delivery of goods and services, the intent is to adjust or amend the policy process and indirectly alter the behaviour of actors involved in policy-making) (Howlett 2000; 2010).

Given the distinct functions and characteristics of OC, VLM and TC crowdsourcing, they can play different roles as policy tools. Arguably, each of the principal types of crowdsourcing can also play various roles. For instance, OC crowdsourcing can be used for surveys, information collection and release, and advertising, and is thus considered as an information/nodality-based tool that can act as an effector or as a defector. Alternatively, it can be used for the community and voluntary organisation of crowds and be considered an organisation-based tool that can be used as an effector for community support or suppression or detector for statistics. However, although increasing, the use of crowdsourcing in the policy cycle has thus far has been limited. Scholars have mainly used OC platforms at the agenda-setting, problem definition and policy evaluation stages; with few exceptions, other approaches such as TC or VLMs have been neglected (Prpić, Taeihagh and Melton, 2015).

*Table 2 Example of policy instruments by principal governing resources (Howlett, Ramesh and Perl [1995], based on Hood (1986)]*

| **Nodality/Information** | **Authority** | **Treasure** | **Organization** |
|---|---|---|---|
| Information collection and release | Command and control regulation | Grants and loans | Direct provision of goods and services and public enterprises |
| Advice and exhortation | Self-regulation | User charges | Use of family, community, and voluntary organizations |
| Advertising | Standard setting and delegated regulation | Taxes and tax expenditures | Market creation |
| Commissions and inquiries | Advisory committees and consultations | Interest group creation and funding | Government reorganization |

It has been suggested that Hood's model (1986) is no longer applicable to 21st century tools such as crowdsourcing (Dutil, 2015), but, as Lendonvirta and Bright (2015) point out, the use of these tools does not replace participatory approaches already in place. On the contrary, it augments them, given the enabling power of the new digital technology.

The speed and ease with which these participations are happening have increased significantly, which in turn results in orders of magnitude increase in the number of participations, decrease the cost of participation, and consequently increased access to dispersed knowledge of the crowds as well as enable challenging power when the best interests of citizen are not taken into account. It must be pointed out that not all applications of crowdsourcing have been with the aim of increasing citizen participation and empowerment in various stages of the policy cycle. Research by Asmolov (2015) and Gruzd and Tsyganova (2015) demonstrates that using volunteers from crowdsourcing platforms is not always benign, and it is possible to prevent collective action by using crowdsourcing as it can be institutionalised (in particular for political purposes), facilitate manipulation (e.g. in the agenda setting process) and decrease transparency (due to the anonymity of certain types of platforms).

At first glance, using the taxonomies of Hood and Howlett, it appears that all of the principal types are substantive in nature, and OC relates to Nodality and Organisation because of dominant thinking about social media (Twitter, Facebook, etc.) and community organisation through voluntary OC platforms (e.g. Enterprise Wikis). Similarly, because of its requiring relatively larger sums of money, TC primarily relates to Treasure and VLMs primarily relate to Organisation.

A closer look, however, reveals that the picture is much more nuanced. In Tables 3 and 4, we highlight the potential for applications of Substantive (Table 3) and Procedural (Table 4) use of VLM, OC, and TC crowdsourcing as policy tools based on the NATO model.

Tables 3 and 4 highlight that the principal types of crowdsourcing can almost be used as every type of policy tool based on the NATO model (1986). Although surprisingly different from the current documented application of crowdsourcing in the literature (Prpić, Taeihagh and Melton, 2015; Liu 2017b), we speculate this is because fundamentally IT-mediated crowdsourcing platforms act as technological enablers and catalysts for the participation of crowds in the policy cycle, and as such can have almost limitless applications in the policy process.

*Table 3 Potential examples of Substantive applications of VLM, OC, and TC crowdsourcing as policy tools, based on Howlett (2010) (D= Detector)*

| Nodality | Authority | Treasure | Organisation |
|---|---|---|---|
| Commissions and inquiries (OC) (D) | Census-taking consultants (local VLM) (D) | Consultants (VLM) (D) | Market Creation (VLM) |
| Information collection (OC, VLM) (D) | Committees and consultations (OC) (D) | Grants, loans, and tax Expenditure (OC, VLM, TC) | Statistics (OC, VLM) (D) |

| Surveys (OC, VLM) (D) | Standard setting and regulation (OC) | Polling policing (Local VLM) (D) | Use of community and voluntary organisations (OC, VLM, TC) |
|---|---|---|---|
| | | Taxes (VLM, OC) | |

Source: Author

*Table 4 Potential examples of procedural applications of VLM, OC, and TC crowdsourcing as policy tools, based on Howlett (2010) (N= Negative, D= Detector)*

| Nodality | Authority | Treasure | Organisation |
|---|---|---|---|
| Information campaigns and advertising (OC, VLM) | Advisory group creation (OC, VLM) (D) | Interest group creation and funding (VLM, OC) | Evaluations (VLM, TC, OC) (D) |
| Information release and notification (OC) | Banning groups and associations (VLM, OC) (N) | Research funding (VLM, TC) (D) | Hearings (OC) (D) |
| Misleading information, propaganda and censorship (N) (OC, VLM) | Agreements and treaties (OC) | Eliminating funding (VLM, OC) (N) | Information suppression (OC, VLM) (N) |

Source: Author

Table 5 examines these potential roles at different stages of the policy cycle.[6] Here we use the front-end (agenda-setting, problem formulation, and policy formulation) and back-end (policy implementation, enforcement, and evaluations) terminology introduced by Howlett (2009). The most commonly observed use of crowdsourcing as a policy tool in the literature is the use of OC as a substantive front-end nodal tool focused on agenda-setting and policy design stages, followed by back-end nodal OC used for policy evaluations and the front-end Treasure use of TC (Prpić, Taeihagh and Melton 2015). The principal types of crowdsourcing as summarised in Table 5 can, however, be used as enablers of almost every policy tool application according to the NATO model. As such the author argues considering crowdsourcing as a policy tool or a definite means of co-production is questionable, and perhaps crowdsourcing should be considered just as a technological enabler that simply can increase speed and ease of participation. In other words if crowdsourcing enables doing everything perhaps it does nothing by itself and just facilitates the speed of participation through providing an enabling environment, as a platform.

---

[6] Various classification attempts and corresponding models of the policy processes exist, of which perhaps the most popular is the use of sequential interrelated stages as a policy cycle. In this article, based on the efforts of Stone (1988) and Howlett, Ramesh and Perl (1995), the policy cycle is seen as a sequence of steps in which agenda-setting, problem definition, policy design, policy implementation, policy enforcement, and policy evaluations are carried out in an iterative manner (Taeihagh *et al*. 2009).

Moreover, these examples from Table 5 show that, although there are convergences around specific themes in terms of the means used, goals for the use of the principal crowdsourcing types can be completely different.

*Table 5 Categorisation of potential applications of principal types of crowdsourcing (VLM/TC/OC) as policy tools in the policy cycle*

F= Front-end (agenda-setting, problem definition and policy formulation)
B =Back-end (Policy implementation, enforcement and evaluation)
D/E= Detector/Effector; S/P Substantive/Procedural; N/P= Negative/Positive

| Application/Type | VLM | TC | OC | D/E | S/P | N/P | NATO type | (Potential) Examples |
|---|---|---|---|---|---|---|---|---|
| **Advisory group creation** | B | | B | | P | | A | VLM or OC participation in advisory groups |
| **Agreements and treaties** | | | F | | P | | A | Use of OCs for treaty verification |
| **Banning groups and associations** | B | | B | | P | N | A | Identification and banning groups online, or locally using volunteers or paid workers |
| **Census-taking consultants** | F | | | D | S | | A | Hiring local VLM participants for conducting census |
| **Committees and consultations** | | | F | D | S | | A | Use of OCs for forming online committees or receiving submissions for white papers etc. |
| **Standard setting and delegated regulation** | | | F | | S | | A | For example the Finish experiment |
| **Commissions and inquiries** | | | B | D | S | | N | Submissions to parliamentary inquiries |
| **Information campaigns and advertising** | B | | B | | P | | N | Advertising using social media, hiring individuals through VLMS to participate in online (or local) campaigns |
| **Information collection and surveys** | F | F | F | D | S | | N | Conducting surveys using social media or VLM platforms and small TC competitions |
| **Information release and notification** | | | B | | P | | N | Release of information using social media |
| **Misleading information, propaganda and censorship** | B | | B | | P | N | N | Use of VLM and OC for identification and censorship of what is deemed as inappropriate. |

| | | | | | | | |
|---|---|---|---|---|---|---|---|
| **Community and voluntary organisations** | F, B | F, B | F, B | | S | | **O** | Supporting formation and participation in non-profit groups using monetary and nonmonetary means; receiving solutions or evaluations |
| **Evaluations** | B | B | B | D | P | | **O** | Use of social media in OC for receiving crowd feedback, use of VLMS for evaluation of programmes, and development of tournaments for evaluation of particular programmes |
| **Hearings** | | | B | | P | | **O** | Use of social media for collection of evidence and participation of crowds in hearings |
| **Information suppression** | B | | B | | P | N | **O** | Voluntary or paid use of crowds for suppressing information using information obfuscation |
| **Market creation** | B | | | | S | | **O** | Formation of particular forms of online markets that can also have offline functionality |
| **Statistics** | B | | B | D | S | | **O** | Collection of statistical data by encouraging voluntary participation of crowds in OC or paid participation of targeted crowds using VLMS |
| **Consultants** | F | | | D | S | | **T** | Hiring consultants from experts workers (e.g. platforms such as odesk, upwork and topcoder) |
| **Eliminating funding** | B | B | | | P | N | **T** | Eliminating previously funded research through TC and VLM platforms |
| **Grants, loans and tax-expenditure** | B | B | B | | S | | **T** | Tax expenditure for funding individuals directly through markets or competitions (e.g. research groups), or indirectly by providing support for creating of OC platforms |
| **Interest group creation and funding** | F | | F | | P | | **T** | Funding for creating websites for participation around specific topics or hiring individuals to participate in activities relevant to special interests |
| **Poll policing** | B | | | D | S | | **T** | Hiring individuals to monitor polls (local VLM (also categorised as a sharing economy) |
| **Research funding** | B | B | | D | P | | **T** | Funding research for large endeavours through TC platforms or use of expert crowds for conducting research using VLMS (e.g. upwork) |

| | | | | | | |
|---|---|---|---|---|---|---|
| **Taxes** | | B | | B | S | T | Use of volunteers of paid workers for identifying tax evasion (e.g. identifying pools using aerial photos for water consumption usage or appropriate property tax) |

Source: Author

Crowdsourcing in policy design

Given the rapid developments in crowdsourcing, the potential it offers in scale-up of the number of individuals involved and rapid acquisition of data and judgements (particularly if expert crowds are involved) is significant for addressing uncertainties surrounding the policy design and analysis (Taeihagh 2017b). Crowdsourcing can increase the level of citizen engagement in policy-making, which has particularly been limited at the policy formulation phase (Prpić, Taeihagh and Melton, 2015; Aitamurto 2012, 2016b; Certoma et al. 2015).

The results from a recent literature review demonstrate that, at present, the use of crowdsourcing in policy design is extremely limited (Prpić, Taeihagh and Melton, 2015). As such, further development of new theoretical frameworks and experiments for exploring and exploiting the potentials that crowdsourcing offers in addressing policy issues are important. Taeihagh (2017b) proposes the examination of new roles for both expert and non-expert crowds at different stages of the policy cycle, as well as an integrated use of crowdsourcing with decision support systems. At present collection, characterisation and examination of the interactions among a large number of policy instruments are difficult. Underutilised types of crowdsourcing, namely VLMS and TCs, can potentially address some of these challenges. For policy design, in particular, crowdsourcing can potentially be used for the collection and characterisation of different policy instruments, examination of the policy instrument interactions, and evaluation of the proposed and implemented policies (Taeihagh 2017b).

Crowdsourcing provides the ability to scale-up the level of engagement by increasing the number of expert or non-expert participants. As a result, it increases the speed of conducting activities when compared with approaches such as organising workshops or conducting offline surveys, as the popularity of crowdsourcing in its different forms increases over time.[7,8] As TCs become more popular and engage more specialised crowds that are able to address complex tasks, and as platforms are further developed, more can be accomplished using crowdsourcing. In addition, increasing the ease of use and accessibility of crowdsourcing platforms will further facilitate their direct integration with decision support systems through Application Programming Interfaces (API).

**Crowdsourcing: a flash in the pan, or here to stay?**

When it comes to speculation about the future of crowdsourcing, there is no shortage of strong views from both its opponents and its proponents. Opponents do not take crowdsourcing seriously; they dismiss it as a fad, citing incidents such as the naming of the Natural Environment Research Council's (NERC) $290-million research vessel as "Boaty McBoatface", following a crowdsourcing campaign in the UK. NERC ignored the outcome of the campaign and named the vessel RSS Sir David Attenborough instead. The subsequent development of a meme based on the incident and numerous documented tales of such online crowd behaviour are given as examples of why crowdsourcing is not to be trusted (Ellis-Petersen 2016). Another danger of crowdsourcing (particularly regarding OCs and the use of social media) is that it allows anyone to distribute information through campaigns, or even to hijack them (Greengard 2011; Prpić, Taeihagh and Melton, 2015), which can facilitate the dissemination of false information.[9,10] On the other hand, proponents of crowdsourcing see immense potential

---

[7] Even in the case of online surveys, the speed at which a worker can carry out a microtask is much faster than an online survey (Prpić, Taeihagh and Melton, 2014).
[8] Expert crowdsourcing, mainly through competition-based platforms (and future high-skilled VLMs sites once their use becomes more mainstream) and non-expert crowdsourcing through the use of VLMs. OC platforms provide access to both expert and non-expert crowds, but require a more sustained effort in attracting and maintaining them. It is worth nothing recent research by Bonazzi et al. (2017) demonstrates a successful combined engagement of expert and non-expert crowds in scenario planning.
[9] OC platforms, for instance, have amplified unscientific and unsubstantiated claims regarding MMR vaccination, resulting in a significant increase in outbreaks of preventable diseases such as measles in the UK and the US (Perry 2013).
[10] A potential worrying development in case of massive adoption of crowdsourcing (such as in the examples highlighted in Table 5) is the difficulty of upholding oversight and keeping organisations accountable in future, especially if block-chain technology is used as the level of anonymity can increase. Block-chain technology such as bitcoin is not anonymous, but in comparison to traditional means of monetary exchange (in the hands of expert individuals) it has a higher level of anonymity as it does not require sending and receiving personally identifiable information: https://bitcoin.org/en/protect-your-privacy.

benefits in its use, and are certain it is here to stay. There are no shortages of claims that it will revolutionise different areas from information collection, processing and management, decision making, to healthcare and learning (Howe 2008; Greengard 2011; Marcus and Parameswaran 2015; Okolloh 2009; Park *et al.* 2017; Turner *et al.* 2012).[11]

Incidents such as "Boaty McBoatface" suggest there is still a clear deficit in understanding how to engage crowds and avoid failures. The subsequent development of a meme based on "Boaty McBoatface" as "Stealthy McStealthface", "Trainy McTrainface", "Footy McFooty Face" and suchlike, demonstrates the backlash from NERC's neglecting the crowd's choice, and has resulted in a sophisticated reaction in other campaigns and highlighted the importance of meaningful engagements (Ellis-Petersen 2016; Boer 2016; Chappell 2017; Hern 2017). Fortunately, the literature illustrates that, when the public are meaningfully engaged, they feel valued and provide productive input (Aitamurto 2012; Sadat 2014; Landemore 2015). Perhaps one of the best know examples being the "Finish experiment", in which Finland's Ministry of the Environment used crowdsourcing to better understand and seek solutions for problems involving the off-road use of fast snowmobiles and ATVs. After receiving input about the problems, a group of citizens and experts evaluated the solutions together, which ensured that the crowd was engaged throughout the process and guaranteed a quality outcome (Aitamurto 2012; Aitamurto and Landemore 2016b).

While some scholars consider application of crowdsourcing platforms as a pathway to sustainability, others have warned against significant regulatory and governance challenges such as the potential for erosion of accountability and tax, transfer of the risks to consumers and users, creation of division among communities, exploitation of crowds, discrimination against individuals, reduction of pay and job security, that need to be addressed (Codagnone, Abadie and Biagi 2016; Aloisi 2015; Taeihagh 2017a; Liu 2017b). Other challenges are due to organizational resistance (Mazumdar et al. 2017), difficulties of assuring quality outputs with scale up, potential for fraud, and manipulation of the platforms through monetary means, administrative privileges, and malicious attacks. Given these complexities, it is difficult to predict what the future holds for crowdsourcing.

---

[11] As a crude measure at the time of finalising this manuscript in November 2017, 469 papers have the term "crowdsourcing" in the title AND mention the term "revolution" in their text. There are also 16,800 academic papers that mention crowdsourcing AND revolution in their text.

Confident predictions have often been made about the future of various technologies that have later returned to haunt those that have made them. For example, Steve Ballmer's famous statement that there was "no chance that the iPhone is going to get any significant market share" or Thomas Krugman's infamous prediction that "the growth of the Internet will slow drastically, as the flaw in 'Metcalfe's law'–which states that the number of potential connections in a network is proportional to the square of the number of participants–becomes apparent: most people have nothing to say to each other! By 2005 or so, it will become clear that the Internet's impact on the economy has been no greater than the fax machine's." These examples steer the author away from making such predictions (Eichenwald 2012; Hendry and Ericsson 2003, p. 66).

However, certain observations can be made:
- at the moment, engagement of crowds through platforms has been manifested in numerous implementations, termed "crowdsourcing", "citizen science", "citizen sourcing", "collaborative innovation", "community systems", "crowd wisdom", "gamification", "open collaboration", "peer production", "prediction markets", "open innovation", etc. As Prpić and Shukla (2016) point out, further development of generalisable frameworks for studying IT-mediated crowds have the potential to unify the field;
- the application of IT-mediated platforms such as crowdsourcing is undeniably increasing in both developed and developing countries in private and public sectors (Prpić, Taeihagh, and Melton 2015; Hira 2017; Taeihagh 2017a; Liu 2017b); correspondingly Figure 1 demonstrates the exponential increase of academic publications with the term "Crowdsourcing" in the title in both Google scholar and Scopus from 2008 to 2016.
- the research for the development of the Internet started in the 1960s and resulted in ARPANET, which can be considered the primitive form of the Internet, but only in the mid-1990s did the Internet became popular in the west (Salus and Vinton 1995; Berners-Lee *et al*. 2000). The infamous statement by Krugman in 1998 for *Time* magazine's 100-year anniversary (and the claim that it was only made according to the requirements of the assignment)[12] was written by a Nobel laureate more than 35 year after the start of research on the technology. It can be argued that the fate of the Internet is not yet clear, and that we are only at the early stages of its development, given there is an expectation of a significant increase (perhaps orders of magnitude) in connectivity through the Internet of Things (IOT: Nordrum 2016). As such, in authors view, when it comes to the use of IT-mediated technologies that use the Internet, such as crowdsourcing, we have thus not yet scratched the surface.

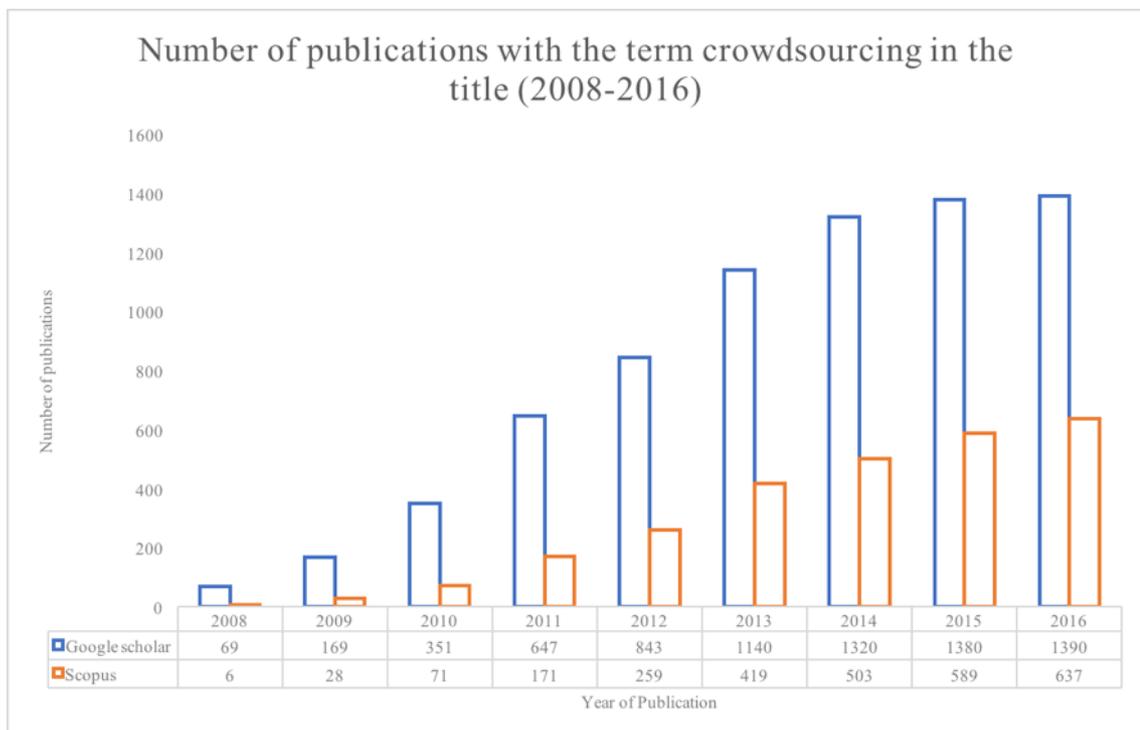

*Figure 1 Number of publications with the term "crowdsourcing" in the title (2008-2016).*

Source: Author

Furthermore, crowdsourcing is rapidly evolving. It is expected that some of the current limitations of crowdsourcing platforms such as inability to use different forms of crowdsourcing simultaneously will be addressed by development of new hybrid crowdsourcing platforms (e.g. expert TC crowds using VLMs or OC.for data collection). There will be further investigation and integration of crowdsourcing with data analysis and machine learning approaches such as in the case of using Natural Language Processing for concept extraction and sentiment analysis from crowdsourced policymaking (Aitamurto *et al.* 2016). Moreover, to date, a vast majority of the research and practice regarding crowdsourcing has been bound to desktop computing, with few mobile applications (Goncalves *et al.* 2015). With continued development of ICT technologies, however, orders of magnitude increase in connectivity, and diffusion of these technologies worldwide, new configurations of software, hardware, and people are emerging (Prpić 2016) that can drastically increase the impact of crowdsourcing, as described below:

- Crowdsensing: also known as participatory sensing or social sensing, enable passive collection of data through the sensors of various mobile devices (for example smartphones) to collect environmental data such as temperature, location, and acceleration, as a consequence of human movement, passively and autonomously sharing the data through wifi/mobile networks through time (Sun *et al.* 2015, Zenonos 2016; Prpić 2016). More specialised data, such as pollution levels, can also be obtained through this technique.
- Situated crowdsourcing: employs IT installations at specific locations to tap into the creativity and problem-solving abilities of crowds. In Hosio's (2016) words, situated crowdsourcing 'simply refers to the process of breaking a large task to smaller pieces, and then offering the subtasks for the public to do using situated technology installations.' Situated crowdsourcing requires more active participation from crowds, where participants use dedicated public installations (such as kiosks and displays) to carry out tasks.
- Spatial crowdsourcing: requires participants to move and carry out tasks at specific locations. In spatial crowdsourcing, researchers explore how to engage crowds to carry out tasks such as taking pictures of signs at specific locations, or undertake tasks relating to emergency response (Krumm and Horvirz 2014; Goodchild and Glennon 2010). Addressing questions such as who can and should be engaged for such tasks, and how much to compensate people for carrying out such tasks, is more complex than traditional VLMS due to the increased complexity of the tasks. Spatial crowdsourcing can also include voluntary services, and overlaps with the sharing economy as with increased focus on use of mobile applications carrying offline tasks become easier[13].

- Wearables crowdsourcing: conducted using embedded sensors in devices attached to the humans through clothing or accessories (Prpić 2016). Wearables crowdsourcing can be used for passive collection and transmission of data about the wearer of the device, or tasks such as monitoring of air or water quality.

---
[12] http://www.businessinsider.com/paul-krugman-responds-to-internet-quote-2013-12/?IR=T.

These new crowdsourcing developments are being adopted mainly in the business domain, but the public sector will also benefit from their use at various stages of the policy cycle, further increasing citizen engagement in the future. For instance, effective introduction of crowdsensing and wearables crowdsourcing can be beneficial in policy enforcement and monitoring, while spatial crowdsourcing can engage citizens in the provision of voluntary services or emergency response. Situated crowdsourcing can be used in agenda-setting or policy evaluation.

As stated at the beginning of this section there are no shortages of strong views on of crowdsourcing from both its opponents and its proponents. Given the complexities of crowdsourcing, its challenges and potentials, as elaborated in this manuscript, such views are to be expected. In its 50th anniversary, Policy Sciences is certainly no stranger to such lively debates, having introduced engaging concepts and techniques such as Wicked

---

[13] Different forms of crowdsourcing and sharing economy share commonalities in terms of the use of reputation systems and IT, the reliance on crowds and the exchange of information and currency (Taeihagh 2017a). The literature in one domain, however, often ignores the other or treats it in a singular form rather than considering the different types that fall under the umbrella term. Sometimes, moreover, a platform is categorised both as a sharing economy and as a crowdsourcing platform by different scholars, particularly when the topic of the study relates to VLMs and OCs (particularly Commons such as Wikipedia). Westerbeek (2016) explicitly differentiates between crowdsourcing and sharing economy platforms by stating the one-on-one, peer-to-peer aspect to be the most important part of a sharing economy, and that this is not present in crowdsourcing. Other scholars distinguish between them by pointing out that, if a labour market platform for instance provides a virtual service that can be performed online (such as Amazon Mturk), that platform is a crowdsourcing platform; in contrast, if it provides a physical service to be performed locally, it is a sharing economy platform (such as TaskRabbit) (Gansky 2010; De Groen, Maselli and Fabo 2016; Aloisi 2015; Rauch and Schleicher 2015). With these new developments in crowdsourcing, however, the line between crowdsourcing and sharing economy platforms seem to be gradually blurring which provides further evidence that as Prpić and Shukla (2016) point out, there is a potential for unifying these fields with development of ganeralisable frameworks for studying IT-mediated crowds.

Problems and Advocacy Coalition Framework that have stayed relevant for decades and still are subjects of scholarship and debate (Rittel and Webber 1973; Sabatier 1988).

Predicting future is difficult, and there are questions about the efficacy of such technologies and their long-term societal consequences, but for now, it is safe to say that given the evidence introduced in this manuscript development and application of IT-mediated technological enablers such as crowdsourcing with all of their challenges and complexities are on the rise.

**Conclusion**

In this article, we briefly introduced the literature on crowdsourcing and considered the three principal types of crowdsourcing, examining their characteristics. We then presented the notion of a generic policy tool, using Hood's NATO model (1986) and Howlett's distinction between substantive and procedural instruments (Howlett 2000, 2010). Using these models, we examined the potential applicability of the principal types of crowdsourcing as different substantive and procedural policy tools, then systematically explored their applications in the policy cycle and highlighted the discrepancy between their current documented use and potential for future use.

By demonstrating the potential for use of crowdsourcing as an enablers of almost every policy tool application according to the NATO model in Table 5 we questioned considering crowdsourcing as a policy tool or a definite means of co-production, and suggested crowdsourcing should be considered just as a technological enabler that simply can increase speed and ease of participation as a platform. We then focused on potential new roles for crowdsourcing at the policy design stage and then discussed the current trends and future trajectories of crowdsourcing.

We hope this article illustrates the new potential uses of crowdsourcing to scholars and practitioners, and that it facilitates the development of more empirical studies (VLMs and TCs in particular) to better understand the efficacy and various potentials for the use of crowdsourcing in the policy cycle as a technological enabler that can increase the speed, ease and rate of participation and as a consequence can reduce costs of participation and increase access to the dispersed knowledge of the crowds.

Platforms on Society, PhD Thesis, TU Delft.

Zenonos, A., Stein, S. and Jennings, N. (2016) An Algorithm to Coordinate Measurements using Stochastic Human Mobility Patterns in Large-Scale Participatory Sensing Settings. Advances in Artificial Intelligence. *Association for the Advancement of Artificial Intelligence (www.aaai.org)*.

Zhang, Y., Gu, Y., Song, L., Pan, M., Dawy, Z. and Han, Z. (2015) Tournament-Based Incentive Mechanism Designs for Mobile Crowdsourcing. *2015 IEEE Global Communications Conference (GLOBECOM)* (pp. 1–6). IEEE.